\documentclass[10pt,conference,compsocconf,letterpaper]{IEEEtran}

\usepackage{todonotes}
\usepackage{amsmath}
\usepackage{subfigure}
\usepackage{multirow}
\usepackage{hyperref}
\usepackage{balance}

\begin{document}
\newtheorem{definition}{Definition}
\newcommand{\nfr}[0]{\textit{nfr}}
\newcommand{\netport}[0]{$\langle\textit{netrange, port}\rangle$}
\newcommand{\fkey}[0]{\ensuremath{f_{\textit{key}}}}

\renewenvironment{description}[1][10pt]
  {\list{}{\labelwidth=0pt \leftmargin=#1
   \let\makelabel\descriptionlabel}}
  {\endlist}

\author{
    \IEEEauthorblockN{Nino Vincenzo Verde\IEEEauthorrefmark{1}, Giuseppe Ateniese\IEEEauthorrefmark{1}, Emanuele Gabrielli\IEEEauthorrefmark{1}, Luigi Vincenzo Mancini\IEEEauthorrefmark{1}, Angelo Spognardi\IEEEauthorrefmark{2}}
    \IEEEauthorblockA{\IEEEauthorrefmark{1}Sapienza - University of Rome, Department of Computer Science, Rome, Italy
    \\\{verde, ateniese, gabrielli, mancini\}@di.uniroma1.it}
    \IEEEauthorblockA{\IEEEauthorrefmark{2}IIT-CNR, Pisa, Italy
    \\\{a.spognardi\}@iit.cnr.it}
}

\title{``No NAT'd User left Behind'': Fingerprinting Users behind NAT from NetFlow Records alone}

\maketitle
\begin{abstract}
  It is generally recognized that the traffic generated by an
  individual connected to a network acts as his biometric signature.
  Several tools exploit this fact to fingerprint and monitor users.
  Often, though, these tools assume to access the entire
  traffic, including IP addresses and payloads. This is not feasible 
  on the grounds that both performance and privacy would be negatively affected.
  In reality, most ISPs
  convert user traffic into NetFlow records for a concise
  representation that does not include, for instance, any
  payloads. More importantly, large and distributed networks are usually
  NAT'd, thus a few IP addresses may be associated to thousands of
  users.
  We devised a new fingerprinting framework that overcomes these
  hurdles. Our system is able to analyze a huge amount of network traffic
  represented as NetFlows, with the intent to track people. It does so by accurately 
  inferring when users are connected to the network and which IP addresses they are using,  
  even though thousands of users are hidden behind NAT. Our prototype
  implementation was deployed and tested within an existing large
  metropolitan WiFi network serving about 200,000 users, with an average load of more than 1,000 users simultaneously connected behind 2 NAT'd IP addresses only.  Our solution
  turned out to be very effective, with an accuracy greater than
  90\%.  We also devised new tools and refined existing ones that may
  be applied to other contexts related to NetFlow analysis.

\end{abstract}




\section{Introduction}

Tracking down individuals on the Internet is fairly simple nowadays. 
Governments collect a huge amount of Internet traffic to carry out mass surveillance programs
and  monitor terrorist organizations or suspects. Analysis and monitoring tools 
exist that can detect and locate anyone by simply looking at Internet traffic logs. 
Indeed, network traffic generated by a single user contains certain patterns that make 
it unique and, thus, discernible. Much work has been done in this area of research and it is now well established that network traffic acts as a biometric signature, or fingerprint, of the user that generated it.
The way it works is that a classifier, based on machine learning, is first trained on the traffic generated by a certain individual to extract and learn distinctive characteristics from it, and then used to trace the same individual from his traffic produced afterwards, while surfing the Internet. 

But, how effective are these classifiers? How realistic is the environment in which they operate?

Classifiers are indeed very effective, but it is often assumed that they are given as input the entire traffic, including payloads, headers, and other timing information. The reality is quite different, however. For performance reasons, the attacker cannot analyze or store the entire traffic, that can reach a throughput higher than $2,000$ Gbit/s in the case of large Internet Exchange points. However, ISPs often convert network traffic into NetFlow records for a more concise representation. These records are used to collect IP traffic statistics for data analysis and contain very little information (no payload, for example). Not all is lost, however, since it may not be difficult for an attacker to devise improved classifiers that can pinpoint individuals by just looking at NetFlow records. Even in this case though, it seems we must at least assume that users are identified by unique IP addresses.  But again reality is harsher, and often users are hidden behind NAT and are all seen outside their network as a single entity, with only one IP address. 

In this paper we show that, quite surprisingly, by mining solely NetFlow data belonging to an Internet Service Provider, or that of an Internet Exchange Point, an attacker is able to track users, and accurately estimate when they are connected to the network and which IP address they are using.  The approach that we propose works also if the target user is hidden behind NAT. This privacy attack corroborates that massive spying activities, such as the ones performed by intelligence agencies (\cite{Brandom2013},\cite{Ehrenfreund2013}) are not only possible, but also require very limited computational effort. 
The fact that it is possible to trace users hidden behind a NAT by just mining NetFlow data is far from being obvious. Indeed, standard classification techniques are ineffective when naively applied in this context. Existing solutions that may work for NAT'd addresses are not designed to work with NetFlow records alone. We devised a new fingerprinting framework that overcomes these hurdles. Our system was tested by targeting users connected through their smartphone to an actually-deployed WiFi network, covering a metropolitan area and its surrounding county, that spans a region of nearly 15,000 $km^2$ and that employs more than 1,300 access points. Such a huge network serves routinely about 200,000 total users, with an average load of more than 1,000 users simultaneously connected behind 2 natted IP addresses only. Despite the ``noise'' of so many users inside NetFlow logs,  users were still fingerprinted with a very high accuracy. In all the analyzed cases, we achieved both {\em precision} and {\em recall} greater than $90\%$ (more on this later). 
Even though we experimented with individuals using smartphones, our approach may be generalized and applied to 
fingerprint users carrying a variety of other devices, such as laptops or tablets.

{\bf Application scenarios.} Some examples of applications of our solutions include:\\
1. {\em After-the-fact forensic analysis.} Since ISPs routinely collect NetFlow records, they might provide assistance to law enforcement agencies and track down criminals after the fact or learn about their habits at different points in time. 
\\
2. {\em Covert intelligence operations.} Officials may be mandated to trace certain individuals on 
national security grounds. While ISPs readily collaborate with their own Governments, they may refuse to provide relevant information to other entities outside the borders. Sometimes even asking for an authorization to access data may not be an option. Our framework can easily be used to monitor the traffic on out-of-bounds routers and fingerprint individuals roaming within the domains of targeted ISPs. 
\\
If abused, our solutions can also be used to violate the privacy of citizens. Therefore, in this respect, our work should be interpreted as a warning sign of what certain organizations might be able to undertake.   

{\bf Contributions.} 
Our main contribution is to provide the first solution for fingerprinting individuals hidden behind a NAT router when only NetFlow records are available.
We believe our solutions are significant since we deployed them within an existing metropolitan WiFi network with thousands of (real) users. 
Along the way, we also devised novel techniques and classification strategies that required appreciable efforts and which we believe are of independent interest (at least to those working with NetFlow analysis). 
In particular, other relevant contributions are:

\noindent
1. A novel method to encode NetFlow records into training sets suitable for several HMM classifiers 
run in parallel. We employed HMMs for their ability to properly capture ``time" information, as in time series analysis, and to handle changes in data distribution over time.\\
2. The definition of a {\em User Detector} that operates on the outputs of multiple HMMs, and
performs a time interval aggregation of the individual results. In our experiments, a simple weighted sum was adopted as aggregation function for its ability to meaningfully combine distinct HMM outputs (more on this later). \\
3. The design of a classification component, based mainly on Random Forest, that can automatically interpret the output of the gating component and supply the final classification.  

{\bf Organization of the paper.}
Section \ref{sec:related} reports on related work.
In Section \ref{sec:backgr-defin}, we introduce the background and the definitions needed to introduce our framework. 
The different components of the framework are detailed in Section \ref{sec:framework}, while Section \ref{sec:experimentsDescription} reports several experimental results that show the viability of our approach. Finally, Section \ref{sec:conclusion} provides the conclusions.

\section{Related work}
\label{sec:related}
Our main claim in this paper is that there is no effective technique to fingerprint individuals
hidden behind NAT when only NetFlow records are available. To justify this claim, we performed
an extensive and rigorous analysis of previous proposals in the area and classified them 
based on (1) the intended target (whether device, host, user, application, etc.) and (2) the technique 
used to perform the fingerprinting/profiling. 
Indeed, it is important to remark that there are many proposed solutions that work for, e.g., devices or applications but cannot be used for individuals. Or, rather, they work for individuals but only when payloads are available (but fail when applied to NetFlow logs only).  

We report the results
of our classification in Table~\ref{tab:related}. We identified three
main strands: works based on \emph{Netflows analysis}, those
based on \emph{stream statistic analysis}, and those based on the
\emph{payload analysis}. With respect to the \emph{target}, we
identified four different categories: \emph{user and host
  profiling}, \emph{traffic and application profiling}, \emph{information leaks} and 
  \emph{device profiling}. 
 

\begin{table*}
\caption{Classification of the related work}\label{tab:related}
\centering
 {%
\newcommand{\mc}[3]{\multicolumn{#1}{#2}{#3}}
\begin{center}
\begin{tabular}{ll|l|l|l|}\cline{3-5}
 &  & \mc{3}{|c|}{\textbf{Used Technique}}\\\cline{3-5}
 &  & \textbf{Netflow Analysis} & \textbf{Stream Statistics Analysis} & \textbf{Payload Analysis}\\\hline
\multicolumn{1}{|c|}{\multirow{4}{*}{\rotatebox{90}{\textbf{Target}}}} & \textbf{User and Host Profiling} & \cite{Melnikov2010a,McHugh2008} & 
\cite{Xu2005a,Xu2008,Wei2006,Chen2007,Karagiannis2007,Pang2007} & \cite{Yan2009,Herrmann2012} \\\cline{2-5}
\multicolumn{1}{|c|}{} & \textbf{Traffic and Application Profiling} & \cite{Barford:2001,Plonka:2000,Rossi2010} &
\cite{Zhang2011,Nguyen2012,Karagiannis2005,Stober2013} & \cite{Dai2013}\\\cline{2-5}
\multicolumn{1}{|c|}{} & \textbf{Information Leaks} &  & \cite{Liberatore2006,Sun2002,Chen2010,Wright2008} & \\\cline{2-5}
\multicolumn{1}{|c|}{} & \textbf{Device Profiling} &  & \cite{Kohno2005,Franklin2006,Desmond2008} &  \\\hline
 \end{tabular}
 \end{center}
}%

\end{table*}

\paragraph{User and Host Profiling}
We distinguish two additional sub-categories:
behavioral targeting and single user/host profiling.  We anticipate, however, that behavioral targeting
 focuses only on identifying communication patterns that
are in common to many users (or hosts), with the intent of 
analyzing anomalies. Thus, it does not consider the same problem addressed 
in this paper.

Behavioral targeting refers to a range of technologies and techniques
used to model the behavior of sets of users or hosts with 
the intent of identifying (1) common features, or (2) anomalies from the typical 
behavior. The main objective of this line of work is to build user profiles 
to offer customized services and it is thus valuable to website publishers and 
advertisers. 
As described in~\cite{Yan2009}, some of these techniques analyze the 
interactions  between users and one or more federated content provider servers. 
Identifying groups of Internet hosts with a similar behavior is convenient 
to detect security breaches, such as DDoS, worms, viruses,
botnets, etc. In~\cite{Xu2005a} and \cite{Xu2008}, the authors profiled 
the Internet backbone traffic with the intent of discovering compromised hosts. 
They analyzed the communication patterns
of end-hosts and services via data mining and
information-theoretic techniques. In the end, they showed how to identify common
traffic profiles, as well as anomalous behavior patterns, that are of
interest to network operators and security analysts. A similar problem
was addressed in~\cite{Wei2006}, where host profiles were used to detect 
anomalous behaviors during the Slammer worm spread. 
We emphasize that our target is different:  we are not interested in creating
a cluster of similar users but rather our focus is to improve our ability
to single out users.

Melnikov et al.~\cite{Melnikov2010a} introduced a proof-of-concept
technique used to distinguish between users by analyzing their NetFlow traffic. 
However, details on how their technique can be
employed in a real-world environment are missing and their 
experiments provide very limited insights. 
In~\cite{Chen2007}, the authors show how to distinguish distinct users
during their online playing activities. However, their approach is not
generalizable and hence not applicable to our context. 
Profiling end-host systems based on their transport-layer behavior 
was proposed in \cite{Karagiannis2007}. There, the authors used graphlets to capture
information flows and inter-flow dependencies. Their basic technique though cannot
be used to re-identify users later on. Re-identification is explicitly considered 
in~\cite{Herrmann2012}, where patterns mined from web traffic are 
used to link multiple sessions of the same user. Unfortunately, this solution 
does not work for users hidden behind NAT.  The same applies for the 
work in~\cite{McHugh2008}, where NetFlows are used to detect behavioral
changes, identify anomalies and trace the propagation of
malware. 
In~\cite{Pang2007}, Pang et al. showed that users can be tracked 
through implicit identifiers in 802.11 networks even when
unique addresses and names are removed. However, their technique
works only when the attacker is physically close to the target and can
eavesdrop the wireless traffic.
A completely different approach is proposed in~\cite{Trestian2008},
where the authors use Open Source Intelligence to improve the results
of host profiling. This refinement can also be used within our framework to 
improve our results. 


\paragraph{Traffic and Application Profiling}
This category aims at recognizing either the type
of network traffic (i.e., p2p, streaming video, VoIP, etc.), or the
application that generated it (i.e., Skype, Gnutella, etc.).
It is a classification technique employed by network
administrators to monitor network traffic and identify different
applications. As shown in Table~\ref{tab:related}, several works
rely on statistics extracted from the network traffic. 
To compute these statistics, it is sometimes sufficient to access just 
the header of TCP/IP packets~\cite{Zhang2011,Nguyen2012,Karagiannis2005,Stober2013}, but
in general the entire packet payload is necessary~\cite{Dai2013}.
These approaches do not scale for large networks where NetFlow analysis is 
preferred. Indeed, in  \cite{Barford:2001,Plonka:2000} and \cite{Rossi2010}, the authors use 
NetFlow analysis for traffic monitoring and application classification. However, their 
techniques cannot  be applied to fingerprint users hidden behind a NAT router. 

\paragraph{Information Leaks}
\emph{Information leaks} category aims at analyzing {\bf encrypted} traffic
with the intent to uncover any useful information.

In~\cite{Liberatore2006} and \cite{Sun2002}, two attacks against
encrypted HTTP streams were presented. The authors identified webpages visited 
by victims based on a pre-built database containing webpage
fingerprints. Chen et al.\cite{Chen2010} make use of fingerprints of
different webpages to infer users' browsing habits. 
Similar approaches have been used to identify spoken
phrases within a VOIP call. In~\cite{Wright2008}, for example, the
authors show how the lengths of encrypted VoIP packets can be used to
identify spoken phrases of a variable bit rate encoded call.

\paragraph{Device Fingerprinting}
\emph{Device fingerprinting} aims at identifying different physical
devices (such as mobile phones).  Microscopic deviations in hardware components, 
such as the clock skews, were used in~\cite{Kohno2005} to fingerprint computer
devices. This technique works even when the device is
behind a NAT or a firewall, and also when the device's system time is
maintained via NTP or SNTP.  Clearly, it won't work to distinguish two 
devices of the same model (i.e., with the same hardware specs). 
The same issue applies to the work in~\cite{Franklin2006}, where a technique to accurately 
identify the software driver used by 802.11 wireless adapters is proposed. 
In~\cite{Desmond2008}, the authors propose to identify devices by using timing analysis 
of probe request frames emitted from wireless client stations when scanning for access
points. However, this approach cannot be used at the network level since wireless frames can 
only be captured within the wireless transmission range.

\section{Background and Definitions}\label{sec:backgr-defin}
This section summarizes two main components used in this work: 
the NetFlow technology and our machine learning approach based on HMMs. 
Netflows are used as building blocks to enable the analysis of the network traffic. 
HMMs are adopted to model the traffic of a user and to re-identify him 
later on, when his traffic will be mixed with the traffic of a number of other users.

\subsection{NetFlows}\label{sec:netflows}
NetFlow is a protocol designed by Cisco to collect IP traffic information, 
getting rid of any IP packet payload. It  makes use of compact
representations of the packet exchange between two network peers. 
NetFlow collects traffic information but discards IP packet payloads
and it is commonly used for network traffic monitoring and reporting.
NetFlow v9 has evolved as a IETF standard called IPFIX, already implemented 
by most network equipment vendors~\cite{B.Claise2008}. 

We decided to focus on NetFlow since most routers natively support it 
and it is the de facto standard for a compact representation of a large
amount of network traffic.  Furthermore, NetFlow records do not 
contain packet payloads, thus the activity of collecting and analyzing them 
is considered legitimate and does not raise privacy concerns (unlike deep packet
inspection). 

A NetFlow enabled device (a \textit{probe}) extracts from each
packet a key composed of specific IP header fields. To simplify the
exposition, we can think to this key as a 5-tuple containing IP
source and destination addresses, source and
destination ports, and the protocol used. More formally, we define a NetFlow 
key function that takes as input an IP packet and outputs a
5-tuple of attributes.
\begin{definition}
  A \textbf{NetFlow key function} \textbf{\fkey} is defined as
  $f_{\textit{key}}: \mathcal{I} \rightarrow \mathcal{K}$, where the
  set $\mathcal{I}$ denotes the set of possible IP packets, and the
  set $\mathcal{K}$ is the set of 5-tuples of the form:
  ($\mathit{IP}_{\mathit{src}},\ \mathit{port}_{\mathit{src}},\
  \mathit{IP}_{\mathit{dst}},\ \mathit{port}_{\mathit{dst}},\
  \textit{protocol}$)\footnote{\small NetFlow v5 includes two more
    fields within the key but they are not relevant to our work. These
    are: the type of service and the index of the ingress
    interface. Newer NetFlow versions have also introduced the
    possibility to specify custom keys and attributes.}
\end{definition}
A NetFlow probe applies the NetFlow key
function to each single packet and dynamically builds in its cache
memory a set of \emph{NetFlow raw records}. These also contain several other attributes, among which
the most relevant are: cumulative number of exchanged packets, bytes counters, flow starting and finishing
timestamps, TCP flags, and Type of Service (ToS). 
More formally:
\begin{definition}
  A \textbf{NetFlow raw record \nfr} is composed by a key value
  $k\in \mathcal{K}$ and a data tuple (\textit{packets},
  \textit{bytes}, \textit{start\_timestamp}, \textit{end\_timestamp},
  \textit{TCP\_flags}, \textit{ToS}). Each element of the data tuple represents a
  feature of the set of IP packets $I\in \mathcal{I}$, with
  $k=\fkey(I)$, exchanged among a single connection between two network peers.
\end{definition}

NetFlow raw records, along with the output of the NetFlow key function,
are subsequently sent via UDP to a NetFlow collector, for storing and
analysis purposes. A new \nfr\ is sent to the collector when the connection is
closed (i.e., a packet explicitly terminates the flow via TCP FIN or RST) 
or the NetFlow expires. Indeed, a NetFlow can expire for three main reasons: 
(1) the flow has been inactive for a time period longer than the
  \textit{inactive timeout};
(2) the flow has been active for a time period longer than the
  \textit{active timeout};
(3) the flow cache is full and some space needs to be freed for new
  flows.
Default values for inactive timeout and active timeout are set to 15 seconds
and 30 minutes, respectively. 

The NetFlow collector may store multiple records for each 
NetFlow key. Indeed, the router may receive a packet with 
the same \fkey\  of an expired NetFlow raw record. In this case, a new 
 \nfr\ with the same key is created.
We will leverage this feature to build our framework later. First, we need
to define the  \emph{flow} as the composition of several related NetFlow 
raw records.
\begin{definition}
\label{def:flow}
  A \textbf{flow} $f$ is defined as a set of NetFlow raw records
  $\{\textit{nfr}_1, \ldots, \textit{nfr}_n\}$ such that $\forall
  1\leq i,j\leq n$, the key associated with $\textit{nfr}_i$ is equal to the key associated with $\textit{nfr}_j$.
\end{definition}
We realized that the flow defined above is very convenient and effective in identifying 
users. For instance, a flow properly captures certain usage patterns and is 
oblivious to NAT routers. Indeed, two NAT'd users connecting to the
same IP address and port will be assigned two distinct local ports. Therefore, 
two distinct flows will be generated, one per each user.  
 

In the following, we define a \emph{bi-directional flow} as
the union of two distinct flows:
\begin{definition}
  Given an IP protocol $\textit{protocol}$, two pairs of IP addresses
  and ports $\textit{ip}_1\textit{:port}_1$ and
  $\textit{ip}_2\textit{:port}_2$, we define a \textbf{bi-directional
    flow} as the union of the flow from
  $\textit{ip}_1\textit{:port}_1$ to $\textit{ip}_2\textit{:port}_2$
  with the one from $\textit{ip}_2\textit{:port}_2$ to
  $\textit{ip}_1\textit{:port}_1$.
\end{definition}
Finally, we will consider only \emph{ordered flow}s, applying a
ordering function \textit{sort} that rearranges a flow, sorting its
\nfr's with respect to the start timestamp. More precisely, we
say that an \textbf{ordered flow} is the flow obtained applying the
\textit{sort} function to the \nfr's that compose the input flow,
namely $\textit{sort}(f)=\{\textit{nfr}_1,\ \ldots,\
\textit{nfr}_n\}$, such that $\forall i<j$,
$\textit{nfr}_i.\textit{start\_timestamp}\leq\textit{nfr}_j.\textit{start\_timestamp}$.

Similarly, we can say that an \textbf{Ordered Bi-directional Flow
  (OBF)} is the bi-directional flow obtained by applying the sort function.
OBFs are therefore sequences of NetFlow raw records 
that describe how the connection between two endpoints evolved over time.
OBFs are essential components in our framework. When properly encoded,
they are used to train HMM classifiers which are described next.

Table \ref{tab:toyexample} reports an example of generation of OBFs. 
Five Netflow raw records are listed as $\textit{nfr}_1$ to $\textit{nfr}_5$. 
Each one is composed
by a key and a data tuple. The key contains source and destination IP addresses, ports, and protocol. 
The data tuple associated with the key reports the number of exchanged packets, bytes, the start and end timestamp, TCP flags and the type of service. Since $\textit{nfr}_1$ and $\textit{nfr}_3$ share the same key value, they are aggregated together in the same flow. As such, the five netflow raw records compose four different flows $f_1$,$f_2$,$f_3$,$f_4$.
Netflow raw records $\textit{nfr}_1$, $\textit{nfr}_2$ and $\textit{nfr}_3$ compose an Ordered Bidirectional Flow. 
Indeed, they are related to the same TCP connection between the IP addresses $192.168.1.10$ and $173.124.18.52$. 
Netflow raw records $\textit{nfr}_5$, $\textit{nfr}_4$ compose another Ordered Bidirectional Flow. Note that $\textit{nfr}_5$ precedes $\textit{nfr}_4$ since the start timestamp of $\textit{nfr}_5$ is prior to the start timestamp of $\textit{nfr}_4$.

\begin{table*}
\caption{Ordered Bidirectional Flows: a simple example}\label{tab:toyexample}
\begin{tabular}{|l|l|l|}
\hline
\textbf{Object} & \textbf{Name} & \textbf{Content}\\\hline
\multirow{4}{*}{\rotatebox{90}{\parbox{1.3cm}{\bf \center Netflow\\ Raw\\ Records}}} & $\textit{nfr}_1$ & key: $(192.168.1.10,5430,173.124.18.52,80,\text{TCP})$, data tuple: $(145,1815,1375690161,1375699541,\text{SIN+ACK},0)$\\
 & $\textit{nfr}_2$ & key: $(173.124.18.52,80,192.168.1.10,5430,\text{TCP})$, data tuple: $(5,421,1375690290,1375699650,\text{SIN+ACK},0)$\\
 & $\textit{nfr}_3$ & key: $(192.168.1.10,5430,173.124.18.52,80,\text{TCP})$, data tuple: $(12,1815,1375690690,1375699703,\text{SIN+ACK+RST},0)$\\
 & $\textit{nfr}_4$ & key: $(192.168.1.10,2345,64.12.121.12,443,\text{UDP})$, data tuple: $(1,196,1375690600,1375699705,\text{SIN+ACK},0)$\\
 & $\textit{nfr}_5$ & key: $(64.12.121.12,443,192.168.1.10,2345,\text{UDP})$, data tuple: $(1,12,1375690590,1375699596,\text{FIN},0)$\\[0.5ex]\hline
\multirow{4}{*}{\rotatebox{90}{\parbox{1cm}{\bf \center Flows}}} & $f_1$ & $\{\textit{nfr}_1,\textit{nfr}_3\}$\\
 & $f_2$ & $\{\textit{nfr}_2\}$\\
 & $f_3$ & $\{\textit{nfr}_4\}$\\
 & $f_4$ & $\{\textit{nfr}_5\}$\\[0.5ex]\hline
\multirow{2}{*}{\rotatebox{90}{\parbox{0.67cm}{\bf \center OBFs}}} & $\textit{OBF}_1$ & $(\textit{nfr}_1,\textit{nfr}_2,\textit{nfr}_3)$\\[0.5ex]
 & $\textit{OBF}_2$ & $(\textit{nfr}_5,\textit{nfr}_4)$ \\[0.5ex]\hline
\end{tabular}
\end{table*}

\subsection{Hidden Markov Models}\label{sec:hidden-markov-models}
 
An HMM is a Finite State Machine able to model a doubly stochastic
process with an underlying stochastic process that is not observable
(it is hidden), but can only be observed through another set of
stochastic processes that produce a sequence of observed symbols, or
vectors, as in our case \cite{rabiner1989tutorial}. We consider HMMs
with continuous observation densities. In its compact form, an HMM is
defined by $\lambda=(A,B,\pi)$:
\begin{description}
\item ${\bf A}$: $N\times N$ state transition probability matrix,
  where $N$ is the number of hidden states. Each matrix element
  $a_{ij}$ is the probability of a transition of the hidden process
  from state $i$ to $j$.
\item ${\bf B}$: $N$ observation probability densities. Each
  probability density represents the probability of a certain
  observable, when the hidden process is in state $i$.
\item ${\bf\pi}$: $N$ length initial state probability vector. Each
  vector element $\pi_{i}$ is the probability that the hidden process
  starts from state $i$.
\end{description}
The use of HMMs is proposed when solving one or more of the following
problems related to the modeled phenomenon \cite{rabiner1989tutorial}:
\begin{description}
\item[Problem 1:] 
  Given an HMM $\lambda$ and a sequence $O$ of observables $o_{1},
  o_{2}, ..., o_{t}$, find the probability P($O$|$\lambda$) that
  these observables are generated by the given model.
\item[Problem 2:] Given a model $\lambda$ and a sequence $O$ of
  observables $o_{1}, o_{2}, ..., o_{t}$, find the sequence $Q$ of
  states $q_{1}, q_{2}, ..., q_{s}$ that maximizes the probability
  P($O$|$\lambda$).
\item[Problem 3:] Given $N$ the number of states, the initial state
  probability distribution $\pi$, a sequence $O$ of observables
  $o_{1}, o_{2}, ..., o_{i}$, and (if known) the corresponding
  sequence $Q$ of states $q_{1}, q_{2}, ..., q_{i}$ that emitted them,
  find the model $\lambda$ that maximizes P($O$|$\lambda$).
\end{description}
To solve the third problem, an iterative procedure is used, that
learns the model adjusting its parameters and optimally adapting them
to the observed training data. The training process is able to create
the best model for the observed phenomenon. The learning is said
supervised when the model can be trained with the knowledge of
both the emitting states and the observables, otherwise it is said
unsupervised.
%

Hidden Markov Models (HMMs) are widely used in
sequences analysis since there exist efficient algorithms to solve the three problems defined 
above~\cite{Durbin98biologicalsequence,Jurafsky,sphinx}.  In this paper, HMMs
are employed to model and recognize user traffic. In this case, the observables are the NetFlow raw
records: as described in Section~\ref{sec:netflows}, each \nfr\ is
represented by a data tuple of $t$ values, corresponding to its
$t$ attributes (number of packets, start timestamps, etc.). In
particular, we consider that the observables are $t$ dimensional
vectors distributed according to $N$ multivariate Gaussian
distributions, one for each state: we will adopt one $N\times t$
matrix, containing the means, and $N$ covariance matrices, to define
the $t$-dimensional multivariate Gaussian distributions. In other words, for each
state of the model, we have $t$ Gaussian densities, one for each of the
$t$ vector elements and, to represents such densities, we have to
specify their means and their covariances.

To realize our framework, we will make use of HMMs to learn the network traffic 
profile of a target user. The training phase will be carried out 
by solving an instance of the Problem 3 above, via an unsupervised 
approach. Then, in the classification phase (when users are subsequently recognized), 
trained HMMs will essentially solve instances of the Problem 1 above. 

\section{The Fingerprinting Framework}\label{sec:profiling-framework}
\label{sec:framework}

\begin{figure*}[ht]
  \centering \hfill
  \subfigure[Overview\label{fig:framework}]{
    \label{fig:dp1}
	\includegraphics[height=0.19\textheight]{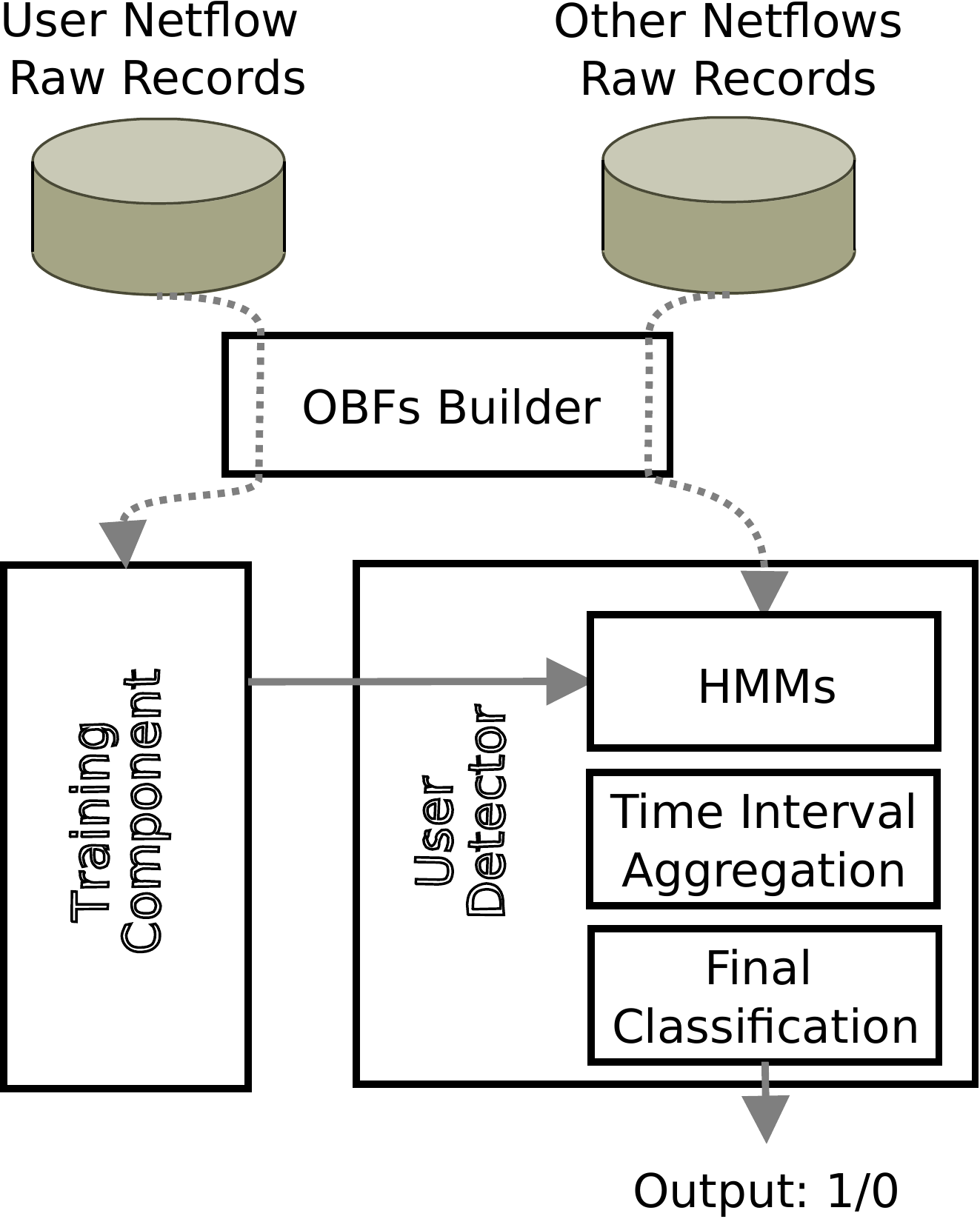}
      }\hfill
  \subfigure[Detail of the Training Component\label{fig:trainingComponent}]{
        \label{fig:dp2}
	\includegraphics[height=0.31\textheight]{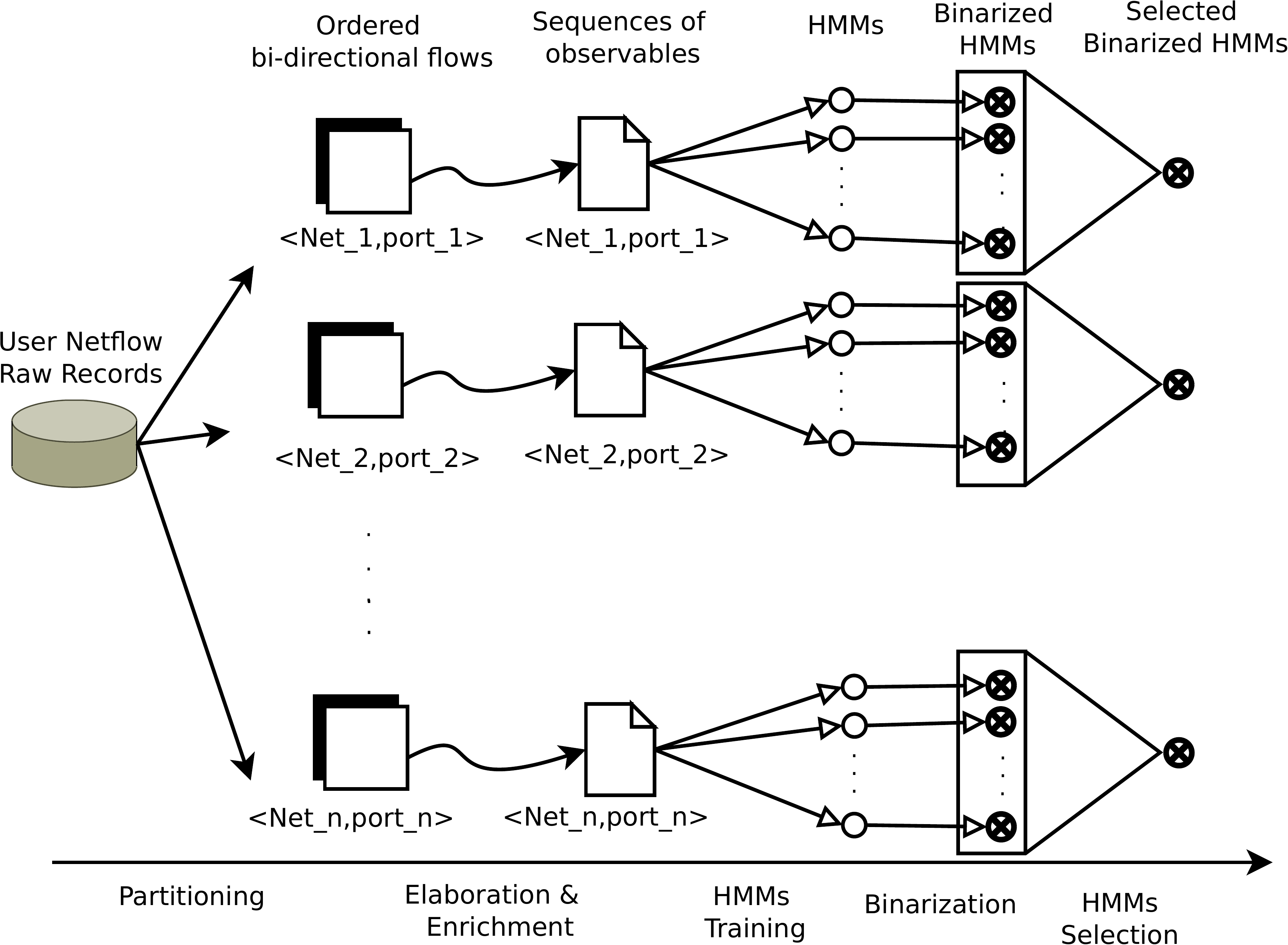}
      }\hfill
      \caption{Our Fingerprinting Framework.}
\end{figure*}

%

Our proposed fingerprinting framework has two main components: the
Training Component, and the User Detector. 
The first component operates as follows: (1) It takes as input NetFlow raw records of the target
user, (2) trains a set of HMMs to recognize its OBFs, and (3) selects the HMMs
that achieve the best performance. 
The User Detector operates as follows: (1) It uses the 
selected HMMs to classify unknown traffic, (2) aggregates the results into a
new dataset that describes time intervals, and (3) applies a final classification
to the aggregated dataset. At the end of this process, the user detector
will determine whether, during a time interval,  the network traffic contains
anything from the target user. 
More details on these two components are provided next.

\subsection{Training Component}
This component is tasked with creating a set of HMMs,
collectively able to recognize the traffic of the target user.  We 
adopted an approach to exploit multiple learners, called mixture of
experts (ME)~\cite{Zhou2012}. Unlike typical ensemble
methods (where individual learners are trained for the same problem), 
a mixture of experts works in a divide-and-conqueror strategy, where a
complex task is broken up into several simpler and smaller subtasks
on which individual learners (the \textit{experts}) are trained. 

In our case, we use a natural task repartition
since the experts are several specialized HMMs, each dedicated to
recognize user traffic towards a single network service.
%
%
Figure~\ref{fig:trainingComponent} graphically describes the process
of realizing trained HMMs in our framework. The entire process is divided into five
sub-phases: Partitioning, Elaboration and Enrichment, HMMs Training,
Binarization, HMM Selection.

\paragraph{Partitioning phase}
We start with a collection of NetFlow raw records  \nfr s that are
generated by user $U$.
In the partitioning phase, these \nfr s
are divided and organized in subsets. 
Each subset is related to a service or to a
set of services accessed by $U$. As in~\cite{Pang2007}, 
we assume that the IP addresses
contacted by the user $U$, along with the corresponding ports, are
\textit{implicit identifiers} of the accessed services. 
However, we must take into account that IP addresses 
may vary at each service request. For example, it is very likely 
to contact two distinct IP addresses when accessing
\textit{www.youtube.com} twice even in a short time interval. 
To overcome this problem, we use the \textit{whois} lookup
protocol~\cite{Daigle2004} to map each IP address to a
netrange that identifies the IP block of addresses that the service
provider controls.  In particular, we select the smallest IP
address range returned by querying five Regional Internet Registries,
namely \textit{arin, ripe, apnic, iana} and \textit{lacnic}.
Combining the retrieved netrange with the contacted port, we obtained a
pair \netport, that constitutes the key used to partition the
collection of the NetFlow raw records of the target user. The last step is the
organization of the \netport\ subsets in OBF sets. This task is carried out 
by the \textit{OBFs builder} that combines and sorts the \nfr's with the same
\netport. This results in several sets of OBFs organized by
\netport. 
Each element of a set represents a sample of the connections between
the target user $U$ and the related service.  It will become an
observation sequence given as input to an HMM. 
 
\paragraph{Elaboration and Enrichment phase}
In this phase, the OBF elements of the above partitioning are
transformed into the observables that will ultimately be used by the HMMs. In
particular, the features of each single \nfr\ of a OBF become the
elements of a feature vector. Therefore, each single OBF becomes a
sequence of feature vector sequences, namely the observation sequences
of the HMMs. The feature vectors are composed of a combination
of the following OBF features:
\begin{description}
\item[Gap:] counts the milliseconds elapsed between a \nfr\ and the
  previous one in the OBF. It is set to $0$ for the first \nfr.
\item[Packets:] is the number of packets reported by a \nfr.
\item[Bytes:] is the number of bytes reported by a \nfr.
\item[Direction:] is the direction of a \nfr. This value is equal to
  $1$ if the related connection was outgoing (originated from the
  target user), 0 if it is ingoing (originated from the other
  end-point).
\end{description}
\paragraph{HMMs Training}
The sequences of feature vectors are  used to train multiple
HMMs in parallel. For each subset of OBFs, we use $80\%$ of
observation vectors to train the HMMs, while we save the remaining $20\%$ for
subsequent phases. Several HMMs are trained by varying the number of 
states and the subset of features of the observation vectors. 
In particular, we used two, three and four states with 14
different combinations of features, namely: \{Pkts\}, \{Bytes\}, \{Gap\},
\{Direction\}, \{Pkts, Bytes\}, \{Pkts, Gap\}, \{Gap, Bytes\},
\{Direction, Bytes\}, \{Direction, Pkts\}, \{Direction, Gap\}, \{Gap,
Bytes, Pkts\}, \{Direction, Bytes, Pkts\}, \{Direction, Gap, Pkts\},
\{Direction, Bytes, Gap\}. These 14 combinations are all the possible 
subsets of cardinality at most 3, that can be achieved starting from 
the set that contains the 4 features:
\{Pkts\}, \{Bytes\}, \{Gap\}, \{Direction\}. 
We set the initial parameters of each HMM
via the K-Means algorithm~\cite{Kanungo2002}. 
The outcome of this phase consists of $42$ uniquely trained HMMs
for each OBF subset created in the previous phases.

\paragraph{Binarization phase}
Each trained HMM is able to evaluate a sequence of feature vectors and
to give the probability that such a new observation was obtained
capturing the traffic of the target user with a given service. This is the
behavior of a probabilistic classifier. The next step is to set a
probability threshold to achieve a binary classifier that recognizes
only two classes: if the observed sequence has a probability lower
than $t$, it will be classified as $0$, otherwise it will be
classified as $1$. 
The threshold $t$ is chosen for each HMM by
testing the $20\%$ of observation vectors set aside during the training
phase, mixed with other observation vectors not belonging to the
target user $U$. In particular,  the threshold is set as the value that
maximizes the accuracy in terms of balanced F-measure $F_1$, namely
\[
F_1=\frac{2PR}{P+R}
\]
where $P$ is the \textit{precision} and $R$ is the \textit{recall}:
precision is the ratio between the positively-and-correctly classified
samples and the positively classified samples, whereas the recall is
the ratio between the positively-and-correctly classified samples and
all the positive samples considered in the test~\cite{Baeza-Yates:1999:MIR:553876}.

%

\paragraph{HMM Selection}
During this last phase, the HMM with the best accuracy is selected. 
Namely, the HMM that maximizes the F-measure is chosen 
among all $42$ possible HMMs available. The corresponding accuracy (in terms of F-measure) will be subsequently used to
assign a weight to the HMM, during the user detection process. 

\subsection{User Detector}
The goal of the Training Component (described previously) is to release a series of
trained HMMs, each of them specialized in recognizing the traffic
of the target user related to a unique network service. The User Detector employs 
these HMMs to analyze some collected traffic and to determine whether there is anything from the target user. 
It aggregates the results by time intervals, and
applies a final 
classification to the aggregated dataset. 

\paragraph{HMM Classification}
The User Detector component starts with a collection of NetFlow raw
records. 
The OBFs Builder is used to combine the \nfr s in OBFs.
Given a new OBFs, the
corresponding HMM is selected, the required features are extracted, and
the result of the classification is stored for further computations.
If there is no specialized HMM for a new OBF, then it  can be discarded as
soon as it arrives. This is because there is no way, in this case,  to determine whether it was 
generated by $U$.
Classification results are stored to be later  retrieved and aggregated with
the ones from the same time interval. In particular, for each single OBF,
we store the predicted class (i.e., $1$ or $0$) and the index of the HMM used 
for the classification.

\paragraph{Time Interval Aggregation}
Once all the OBFs of the time interval have been classified, we use an
aggregation function to summarize the results. The aggregation
will generate a concise record whose length depends on the
number of HMMs trained for the target user $U$. 
The record will represent the weighted number of OBFs that each HMM has
attributed to the user $U$ during the time interval. 
Therefore, a record will be of the form:
\[
\langle\texttt{weight}(\texttt{HMM}_1)\times\texttt{OBF}_{1}(U); \ldots;
\texttt{weight}(\texttt{HMM}_n)\times\texttt{OBF}_{n}(U)\rangle
\]
where $\texttt{weight}(\texttt{HMM}_i)$ is the weight assigned to the
$i^{th}$ HMM (that is its accuracy), $\texttt{OBF}_{i}(U)$ is the number of OBFs recognized
by $\mathtt{HMM}_i$ as belonging to the user $U$ during that interval,
and $n$ is the number of trained HMM.

\paragraph{Final Classification}
The User Detector has to finally associate a binary label to the
record generated after the Time Interval Aggregation: $1$ indicates
that the user was connected to the network during that time period,
$0$ otherwise. A naive solution would be to sum all the values composing a
single record, and fix a threshold to convert the sum into a binary
result. However, better results can be attained by using standard
classification algorithms applied on records from several time
intervals.  In the experiments section, we will compare the
performance of several classification algorithms, such as Support Vector
Machine, Random Forest, JRip, Multilayer Perceptron and Naive Bayes. In the end,
this approach turns out to be very effective,  providing a very high \textit{precision} and \textit{recall}.


\section{Experiments and Discussion}
\label{sec:experimentsDescription}
In this section we briefly describe the implementation of the proposed
fingerprinting framework and how the experiment environments were set up. Then,
we discuss and report on the results we attained.

\subsection{Framework implementation}
To prove the effectiveness of our proposal, we realized a working
prototype of our fingerprinting framework, as designed in
Figure~\ref{fig:framework}. The core system has been coded within a
Java environment.  In particular, both the
training component and the user detector component have been implemented in Java,
and a Java/R Interface called
JRI\footnote{\small \url{http://www.rforge.net/JRI/}} has been used to run R
from the Java application.  Basically, HMMs have been implemented in R, 
with the support of the package
\emph{mhsmm}\footnote{\small \url{http://cran.r-project.org/web/packages/mhsmm/index.html}}.
This package provides parameter estimation and prediction for HMMs for
data with multiple observation sequences, and supports Multivariate
Gaussian distributions.  The Final
Classification phase has been realized by using the Weka library,
that provides a large suite of machine learning algorithms.  In the
following, the experiment environments and the attained results are
discussed in details.

\subsection{Experiment environments}
\label{sec:expenvironments}
The typical hardware configuration to keep track of NetFlow generated
by user traffic is composed of one or more routers and switches with
NetFlow capabilities (the \textit{probes}) and a NetFlow
collector. During its normal duty, each NetFlow-enabled device (router
or switch) creates and elaborates NetFlow records for every single
packet. Then, it transmits  batches of collected
records to the collector that stores them for further
analysis.  
We set up two different environments where we collected NetFlow traces
from real users, to realize both a small scale and a large scale
experiment.

For the small scale experiment, we configured a NetFlow-enabled router
and a single wireless access point to create a WiFi network within our premises. 
The router was the default gateway for the network and was configured to send NetFlow data
to a local collector. 
We profiled a total of $26$ different
users accessing the Internet with their mobile devices during one
month of monitoring, collecting $4.8$ GB of traffic and more than 
500,000 NetFlow raw records.

\begin{figure}	
    \centering    
    \includegraphics[width=0.8\linewidth]{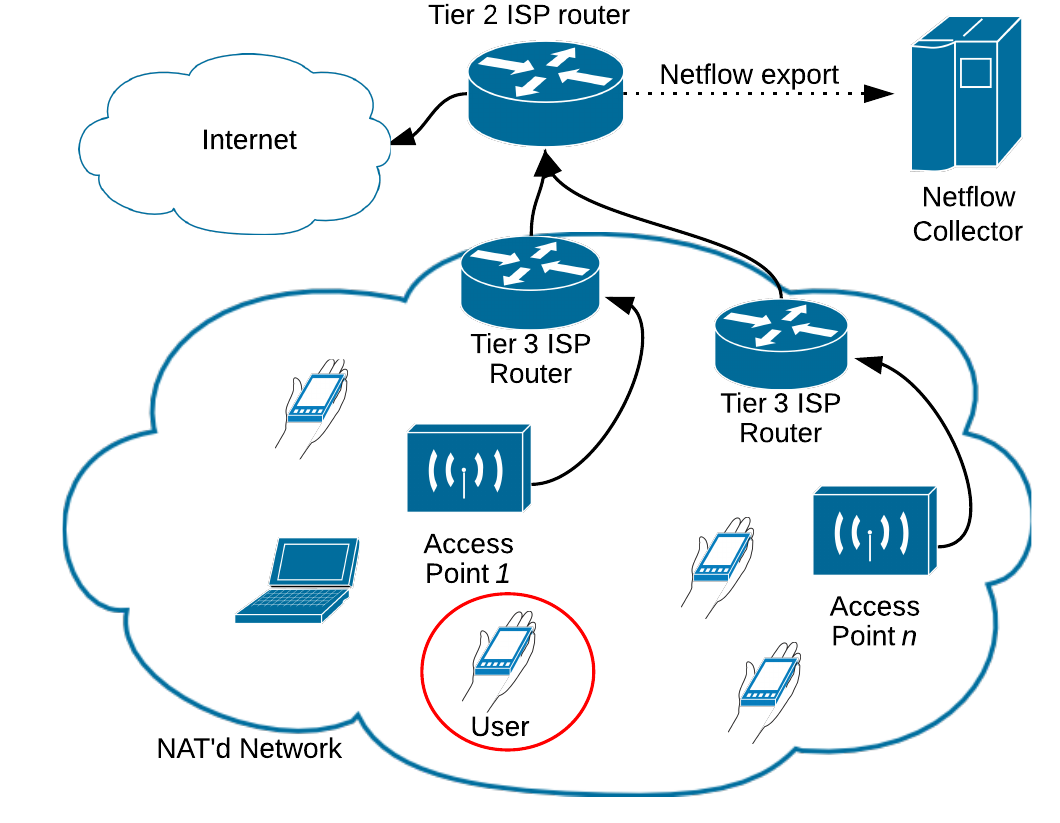}
    \caption{Large Scale Experiment Network}
\label{fig:largescalenetwork}
\end{figure}

For the large scale experiment, we had access to the NetFlow traffic
of a metropolitan network that provides public WiFi
connectivity to a large region with 200,000 registered users, covering 144 towns in an area spanning 
nearly 15,000~$\textit{km}^2$, and counting about 1,300 wireless access points 
(Figure~\ref{fig:largescalenetwork} outlines the network configuration).
A NetFlow probe was placed inside the ISP that provides connectivity services 
for the public WiFi network. 
The public WiFi network is NAT'd and uses only $2$ public IP addresses
with an average of $250$ GB of daily traffic, thus producing nearly $20$ millions
NetFlow raw records per day. 

%

\subsection{Small Scale Experiment}
\label{sec:results}
\begin{figure*}	
    \centering    
    \includegraphics[width=0.7\linewidth]{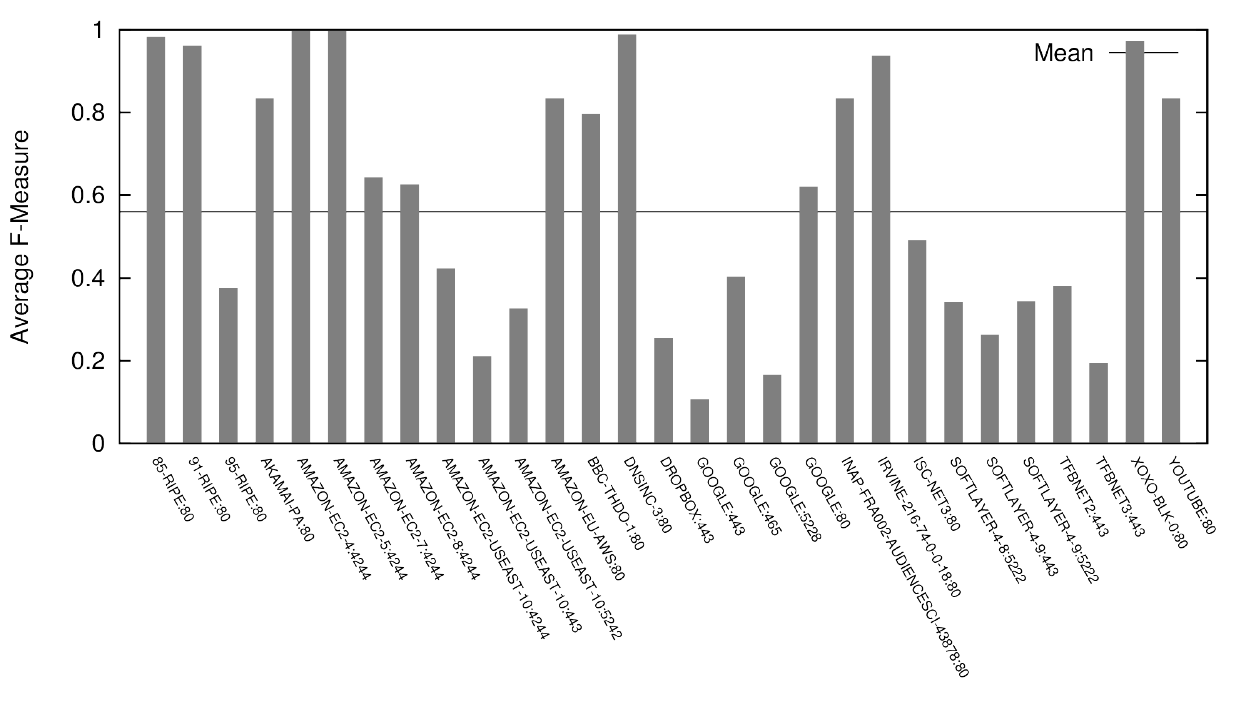}
    \caption{Ordered Bi-directional Flows classification. Accuracy per service for a specific user}
\label{fig:HMMAccuracyByService}
\end{figure*}

We monitored the Internet connection of $26$ different users for one month,
collecting their traffic when they were connected to our access
point. The first week's NetFlow data was used for the Training Component, the rest to train and test the entire framework through cross-fold validation. During the training week, the smartphones of the $26$ target users contacted a total of $74$ different services (composed of netrange,port pairs). 
\begin{figure}[t]	
    \centering    
    \includegraphics[width=.8\linewidth]{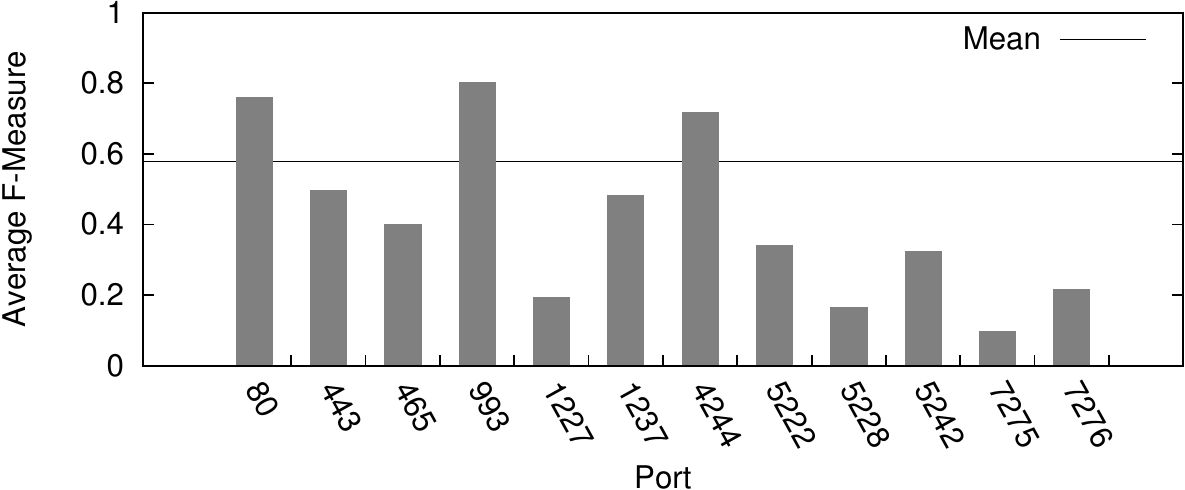}
    \caption{Ordered Bi-directional Flows classification. Average accuracy by port number}
\label{fig:HMMAccuracyByPort}
\end{figure}
Figure~\ref{fig:HMMAccuracyByPort} reports the average accuracy, in terms of F-measure, of the HMMs trained to recognize the traffic toward these services aggregated by port.
It can be seen that HMMs devoted to recognize the traffic toward port 80 have an average accuracy of $0.76$, while those devoted to recognize the traffic toward  port 443 (which is encrypted) have an accuracy of $0.49$. 
It may seem that the encryption reduces the accuracy. However, this is not completely accurate. Indeed, the features that we use to train our classifiers (like the gap or the packets transmitted) are not significantly influenced by the encryption. This is also confirmed from the results achieved analyzing the encrypted traffic flowing on port 993 (IMAP over TLS/SSL). 
Indeed, the HMMs on this port have accuracy equal to $0.80$ in average. 
On the other hand, port 7275 has very low accuracy. This port is related to the Open Mobile Appliance User Plane Location protocol that is used by mobile devices to receive GPS info quickly. Thus, the traffic to this port is somehow 
automatic and this explains the low accuracy we measured. 
In general, certain services are well suited for identifying users  while others are more ``impersonal''. This is confirmed also by Figure~\ref{fig:HMMAccuracyByService}. It details the accuracy of all HMMs employed by the User Detector to recognize the traffic of a specific user (we selected the worst performing). 
Notice that services hosted by Google, and accessed through ports different than $80$, have a fairly low accuracy. On the other hand, the Amazon Elastic Compute Cloud (Amazon EC2), a service used by many application developers, reaches very high accuracy. 
 
Services that have a higher accuracy are the ones that should be used to fingerprint individuals. This is why, in the time interval aggregation, the accuracy is represented by a weight. 
Note also that several services are very popular while others are used by very few users (who are then easier to profile). For instance, the service DNSINC-3 in the figure is related to an Android application called \emph{DynDNS client} from Dynamic Network Services, Inc., and that application counts less than 50,000 downloads worldwide. 

Other than the accuracy of the OBFs classification, we must also measure the overall performance of the framework
to determine when the target user is connected. For this, we tested several  classification algorithms for the final 
classification phase. 
%
%
Table~\ref{tab:rm1Performance} sums up the results that we achieved by using five different algorithms: Random Forest, Naive Bayes, Multilayer Perceptron (MLP), Support Vector Machine (SVM) and JRip (we used the Weka implementation of these algorithms). Random Forest behaved better than the rest in all the evaluation metrics that we considered. It reaches $95\%$ of true positive rate, and only $7\%$ of false positive rate. Precision, Recall and F-measure (i.e., the harmonic mean of precision and recall) are equal to $0.95$, $0.93$, $0.94$, respectively. 

\begin{table}[t]
 \caption{Small scale experiment---Time interval classification. Comparison of five different classification algorithms used in the final classification phase.  Average values are reported. }
 \label{tab:rm1Performance}
 \begin{center}
\begin{tabular}{lrrrrrrrr}
\textbf{Algorithm} &\textbf{TPR} & \textbf{FPR} & \textbf{Prec} & \textbf{Recall} & \textbf{F-measure} \\
Rand. Forest &0.95&	0.07&	0.95&	0.93&	0.94\\
Naive Bayes &	0.55 &	0.08&	0.55&	0.87&	0.67\\
MLP &	0.86&	0.14&	0.86&	0.88&	0.87\\
SVM&	0.69&	0.09&	0.69&	0.88&	0.76\\
JRip&	0.94&	0.12&	0.94&	0.89&	0.91
\end{tabular}
\end{center}
\end{table}

The area under the ROC (Receiver Operating Characteristic) is a convenient way of comparing classifiers. A random classifier has an area of 0.5, the ideal one has an area of 1. Under this metric, we confirmed that Random Forest performs better than other algorithms within our framework (SVM was the worst)~\cite{fawcett04roc}. 
Indeed,  Table~\ref{tab:rm1ClassificatorComparison} reports the average ROC Area related to $6$ different users selected among the $24$ that we profiled (individual values are also reported). It shows that, for all users, Random Forest reaches a ROC area close to 1 (from $0.95$ to $0.97$) and has a low variance as well. 
The latter reveals that Random Forest is appropriate for classifications involving distinct users, with only a small variation over the final performance. 


\begin{table*}
\caption{Small scale experiment---Time interval classification. ROC Area is reported for six targeted users and five  algorithms used in the Final Classification phase. }\label{tab:rm1ClassificatorComparison}
\newcommand{\specialcell}[2][c]{%
\textbf{
  \begin{tabular}[#1]{@{}c@{}}#2\end{tabular}}}
\centering
\begin{center}
\begin{tabular}{lrrrrrrrr}
\specialcell{Classification Algorithm} & \textbf{User 1} & \textbf{User 2} & \textbf{User 3} & \textbf{User 4} & \textbf{User 5} & \textbf{User 6} & \specialcell{ROC\\Mean} & \specialcell{ROC \\Variance}\\
\textbf{JRip}	&0.93	&0.94	&0.88	&0.9	&0.90	&0.90	&0.91 & $4.97\times 10^{-4}$\\
\textbf{SVM}	&0.92	&0.92	&0.8	&0.66	&0.86	&0.63	&0.80 & $161.77\times10^{-4}$\\
\textbf{MLP}	&0.85	&0.95	&0.91	&0.89	&0.93	&0.86	&0.90 & $15.36\times10^{-4}$\\
\textbf{Naive Bayes}	&0.79	&0.92	&0.86	&0.86	&0.93	&0.76	&0.85 & $46.26\times10^{-4}$\\
\textbf{Random Forest}	&\textbf{0.96}	&\textbf{0.96}	&\textbf{0.95}	&\textbf{0.97}	&\textbf{0.96}	&\textbf{0.95}	&\textbf{0.96} & $\mathbf{0.56\times10^{-4}}$
\end{tabular}
\end{center}
\end{table*}


\subsection{Large Scale Experiment}

\begin{table*}
 \caption{Large Scale Experiment: Performance of our fingerprinting framework.}
 \label{tab:ExtrabirePerformance}
 \newcommand{\specialcell}[2][c]{%
\textbf{
  \begin{tabular}[#1]{@{}c@{}}#2\end{tabular}}}
 \begin{center}
\begin{tabular}{lrrrrrrrr}
\specialcell{User} &\textbf{TPRate} & \textbf{FPRate} & \textbf{Precision} & \textbf{Recall} & \textbf{F-measure} & 
\specialcell{Correctly \\Classified Hours}
 & \specialcell{Incorrectly \\Classified Hours}\\
Suspect 1 &1 &	0&	1&	1&	1 & 24 & 0\\
Suspect 2 &	1 &	0.07&	0.91&	1&	0.94 & 23 & 1\\
Suspect 3 &	0.92 &	0.08&	0.92&	0.92&	0.92 & 22 & 2\\
Suspect 4 &	0.9&	0&	1&	0.9&	0.95 & 23 & 1\\
Suspect 5 &	0.9&	0&	0.96&	0.96&	0.96 & 22 & 2 
\end{tabular}
\end{center}
\end{table*}

For the large scale experiment, we used the large WiFi network described in Section \ref{sec:expenvironments}. 
Consider a scenario in which an intelligence agency intents to monitor and trace 
a group of, say, five suspects. The agency, with no wiretap warrant or direct access 
to the large WiFi network, can only export all the NetFlow data produced by the main ISP router providing connectivity to the large WiFi network (see Figure \ref{fig:largescalenetwork}).
For the sake of the experiment, we enrolled five volunteers that used their own mobile phone, without installing new applications or changing their usual behavior. Two of them had a pristine phone with no third-party applications (a worst-case scenario for our profiling framework), but with certain services correctly configured, such as (1) Gmail, (2) Twitter,
(3) Facebook, (4) Skype, in addition to (5) the backup of the phone camera (via Dropbox). 
All five volunteers were previously profiled for a period of $8$ hours by inducing them to connect to an access point directly controlled by us (acting as the intelligence agency). The average training traffic collected per user was of $12.64$ MBytes.

During the test phase, an average of around $1{,}100$ users were connected to the WiFi network simultaneously, and all of them 
were NAT'd behind only two IP addresses. 
During the experiment, we attempted to detect the presence of each one of the suspects by mining the NetFlow data. The suspects used the WiFi network only during specific hours, from 10AM to 10PM.
Approximately $100$ million NetFlow raw records were analyzed during the test phase that lasted $24$ hours. This roughly corresponds to $~96$ GB of traffic, with an average of $9$ Mb per second (Mbs). 
About $700.000$ unique netranges were contacted by the $1{,}100$ users. Only the $0.02\%$ of these netranges were contacted during the training phase by the suspects.


Table~\ref{tab:ExtrabirePerformance} shows the results achieved. Note that we report only on the results 
achieved with the Random Forest classifier since it is the best performer in the final classification step. 
In all cases, the true positive rate is higher than $0.9$, while the false positive rate is lower than $0.08$. Furthermore, {\em precision} and {\em recall} are higher than $0.9$ in all cases. 
\textit{Suspect 1} was successfully detected, while \textit{Suspect 2} and \textit{Suspect 4} produced one false positive and one false negative only. 
\textit{Suspect 3} and \textit{Suspect 5} were misclassified for two hours but these were the users with pristine phones running only applications installed by the phone manufacturer.

\begin{figure*}[ht]
  \centering 
  \subfigure[Suspect 1. ROC Area: 1\label{fig:user1}]{
	
\includegraphics[width=0.18\linewidth]{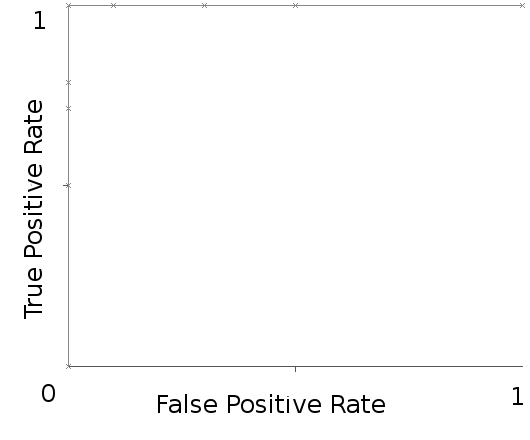}
      }
  \subfigure[Suspect 2. ROC Area: 0.99\label{fig:user2}]{
	\includegraphics[width=0.18\linewidth]{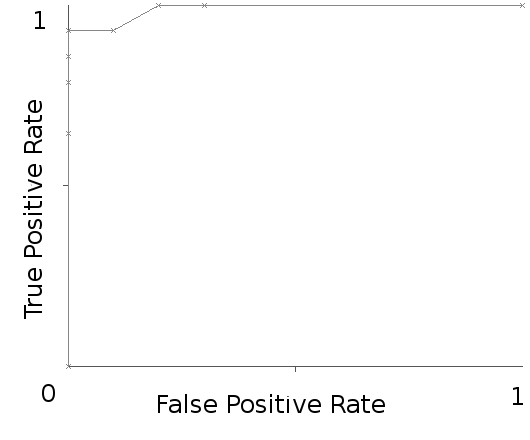}
      }
  \subfigure[Suspect 3. ROC Area: 0.98\label{fig:user3}]{
	\includegraphics[width=0.18\linewidth]{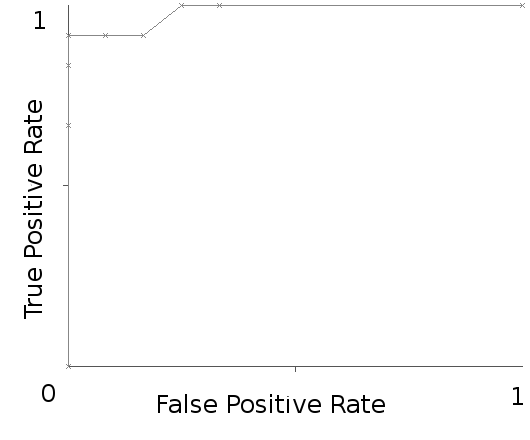}
      }
  \subfigure[Suspect 4. ROC Area: 0.99\label{fig:user4}]{
	\includegraphics[width=0.18\linewidth]{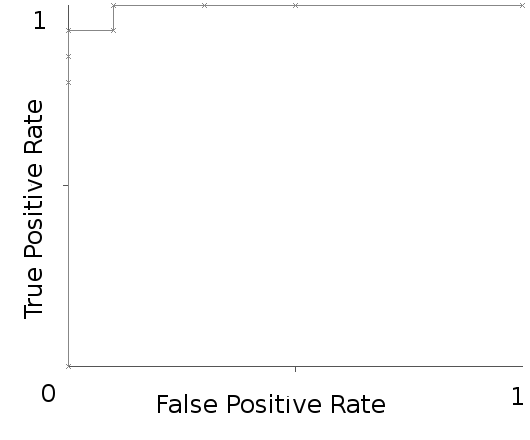}
      }
  \subfigure[Suspect 5. ROC Area: 0.95\label{fig:user5}]{
	\includegraphics[width=0.18\linewidth]{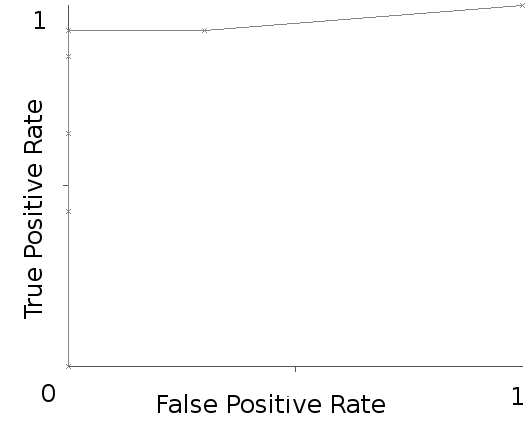}
      }
      \caption{ROC curves for each suspected user.\label{ROC}}
\end{figure*}

Figure~\ref{ROC} shows the ROC curves achieved when fingerprinting the
five target suspects. It can be noticed that in all cases the area
under the ROC curves is greater than 0.95. 

\section{Concluding Remarks}
\label{sec:conclusion}
We showed that it is possible to fingerprint NAT'd individuals when only NetFlow records are available. This is 
a very realistic scenario since most networks are NAT'd and ISPs generally deal only with NetFlow data.
We implemented and tested our framework on an existing metropolitan WiFi network with thousands of real users. 
Our solution uses a series of properly trained HMMs and cleverly combines them to maximize the success rate. 
Along the way, we also devised new tools and refined existing ones that could be generalized and applied to other contexts related to NetFlow analysis.

We are currently working on improving the performance of our fingerprinting framework by leveraging inter-dependencies of distinct network services. Furthermore, we are refining our implementation to be deployed as a 
real-time user localization prototype. Finally, we are also investigating possible countermeasures. It seems solutions such as TOR or VPN tunneling may degrade the performance of our approach, but it is not yet clear whether they can preclude it completely.

\bibliographystyle{abbrv}
\balance
\bibliography{NetflowUserTracking}


\end{document}